\documentclass[twocolumn,showpacs,prc,nofootinbib,floatfix,axodraw]{revtex4}
\usepackage[sort&compress]{natbib}

\usepackage{mathrsfs}
\usepackage{graphicx,booktabs,bm}
\usepackage{overpic}
\usepackage{color}
\usepackage{amssymb}
\usepackage{hyperref}
\usepackage{soul}

\usepackage{mathrsfs,bm,amsmath,amssymb}
\usepackage{longtable,lscape}
\usepackage{txfonts}
\usepackage{amssymb}
\usepackage{indentfirst}
\usepackage{graphicx,booktabs}
\usepackage{multirow,ulem}
\usepackage{subfigure}

\usepackage{float}
\usepackage{array}

\begin{document}
\title{ Nucleon-nucleon interaction in the $^3S_1$-$^3D_1$ coupled channel for a pion mass of 469 MeV}

\author{Qian-Qian Bai}
\affiliation{School of Physics, Beihang University, Beijing 102206, China}

\author{Chun-Xuan Wang}
\affiliation{School of Physics, Beihang University, Beijing 102206, China}

\author{Yang Xiao}
\affiliation{School of Physics, Beihang University, Beijing 102206, China}
\affiliation{Universit\'e Paris-Saclay, CNRS/IN2P3, IJCLab, 91405 Orsay, France}
\affiliation{School of Space and Environment Science, Beihang University, Beijing 102206, China}

\author{Jun-Xu Lu}
\affiliation{School of Space and Environment Science, Beihang University, Beijing 102206, China}
\affiliation{School of Physics, Beihang University, Beijing 102206, China}

\author{Li-Sheng Geng}
\email[Corresponding author: ]{lisheng.geng@buaa.edu.cn}
\affiliation{School of Physics, Beihang University, Beijing 102206, China.}
\affiliation{Beijing Key Laboratory of Advanced Nuclear Materials and Physics, Beihang University, Beijing, 102206, China}
\affiliation{School of Physics and Microelectronics, Zhengzhou University, Zhengzhou, Henan 450001, China}

\begin{abstract}

In this work, we apply  the  relativistic chiral nuclear force to describe the state-of-the-art lattice simulations of the nucleon-nucleon scattering amplitude.  In particular, we focus on the $^3S_1$-$^3D_1$ coupled channel for a pion mass of 469 MeV. We show that at leading order the relativistic chiral nuclear force can only describe $\delta_{3S1}$ and $\varepsilon_1$ up to $T_\mathrm{lab.}\approx10$ MeV, while at the next-to-leading order it can do much better up to $T_\mathrm{lab}=200$ MeV. However, at the next-to-next-to-leading order, the description deteriorates, which can be attributed to the fact that the pion-mass dependence of the pion-nucleon couplings $c_{1,2,3,4}$ may not be negligible. Furthermore, all the studies consistently yield negative $\delta_{3D1}$, contrary to the lattice QCD results which are positive but consistent with zero. The present study is relevant to a better understanding of the lattice QCD nucleon-nucleon force and more general baryon-baryon interactions.
\end{abstract}

\date{\today}


\maketitle

\section{Introduction}
Lattice QCD simulations have become an indispensable tool to explore the non-perturbative baryon-baryon interactions, in particular, hyperon-nucleon and hyperon-hyperon interactions~\cite{Aoki:2020bew,Illa:2020nsi,Wagman:2017tmp,Inoue:2010es,Beane:2010hg,Ishii:2006ec}, which have so far remained difficult to be studied experimentally. Nonetheless, lattice QCD simulations of coupled-channel baryon-baryon interactions are  challenging as well (see, e.g., Refs.~\cite{Briceno:2013bda,Sasaki:2019qnh}). In this sense,  reliable theoretical studies of lattice QCD simulations are in urgent need, particularly for those simulations performed at unphysical pion masses, which remain unconstrained. One of such theoretical tools is chiral perturbation theory~\cite{Weinberg:1978kz,Gasser:1983yg,Gasser:1987rb}(see Ref.~\cite{Scherer:2012xha} for a pedagogical review) and the resulting chiral nuclear forces~\cite{Weinberg:1990rz,Weinberg:1991um,Ordonez:1995rz,Bedaque:2002mn,Epelbaum:2008ga,Machleidt:2011zz}. It should be noted that the conventional chiral nuclear forces are based on the heavy baryon (HB) version of chiral perturbation theory~\cite{Jenkins:1990jv,Bernard:1992qa}.    In recent years, it has been proposed that one can build a covariant version of chiral nuclear forces with the covariant baryon chiral perturbation theory~\cite{Fuchs:2003qc,Geng:2013xn}. Such proposals have been explored mainly by two groups, the Bochum group~\cite{Epelbaum:2012ua,Baru:2019ndr,Ren:2019qow,Ren:2022glg} and the BUAA group~\cite{Ren:2016jna,Li:2016mln,Li:2018tbt,Xiao:2018jot,Song:2018qqm,Xiao:2020ozd,Bai:2020yml,wang:2020myr,Liu:2020uxi}. In a recent work, the covariant chiral nuclear force has been constructed up to the next-to-next-to-leading order~\cite{Lu:2021gsb,Lu:2022yxb}, where an accurate description of the PWA93 phase shifts~\cite{Stoks:1993tb} can be obtained up to $T_\mathrm{lab.}=200$ MeV, comparable to those of the next-to-next-to-next-to-leading-order non-relativistic chiral nuclear forces~\cite{Machleidt:2011zz,Epelbaum:2014sza}.

In Ref.~\cite{Inoue:2011ai}, the HALQCD Collaboration reported lattice QCD simulations of the nucleon-nucleon phase shifts in the $^1S_0$ and $^3S_1$-$^3D_1$ channels for five pion masses, i.e. 1171, 1015, 837, 672, and 469 MeV, which will be refereed to as lQCD1171, lQCD1015, lQCD837, lQCD672, and lQCD469 for the sake of convenience in the remaining part of the present work. These state-of-the-art results not only provide a confirmation of the long accepted qualitative picture of the nuclear force, but also offer valuable constraints on the light-quark dependence of chiral nuclear forces. In Ref.~\cite{Hu:2020djy}, the $^1S_0$ and $^3S_1$ phase shifts were studied in leading order Weinberg chiral effective field theory up to $T_\mathrm{lab}\approx150$ MeV. By fitting to the phase shifts of different pion masses separately with the two LO LECs and the cutoff, they obtained reasonable  descriptions of the lattice QCD data. In Ref.~\cite{Bai:2020yml}, the lattice QCD data were studied in leading order covariant ChEFT adopting a different strategy, namely the lQCD469 data were analyzed together with the physical $NN$ phase shifts, while the larger pion mass data lQCD672, lQCD837, lQCD1015, and lQCD1171 were studied by neglecting the one-pion exchange contributions. The separation of the lattice QCD data into two groups was motivated by the observation that for the large pion masses, the one-pion exchange contributions can be approximated as contact interactions such as the pionless version of the covariant ChEFT is appropriate. The good fits (up to $T_\mathrm{lab.}=80$ MeV) indicated that this was indeed a reasonable treatment.

Nevertheless, we note that the description of the lQCD469 data together with the physical $NN$ phase shifts is a bit tricky. On one hand, the $^1S_0$ phase shifts can be described well with the same four LECs and  cutoff. On the other hand, the description of the $^3S_1$-$^3D_1$ coupled channel is less satisfactory, particularly the $^3D_1$ phase shift $\delta_{3D1}$ and the mixing angle $\varepsilon_1$. More specifically, the mixing angle $\varepsilon_1$  obtained at $m_\pi=469$ MeV is very close to the physical one. Meanwhile, $\delta_{3D1}$ obtained at $m_\pi=469$ MeV is positive (but consistent with zero within uncertainties) while ChEFT seems to prefer  negative $\delta_{3D1}$, similar to the physical case. 

Given  that  both lattice QCD and ChEFT are widely accepted as model independent approaches to deal with the non-perturbative strong interaction in general and the nucleon-nucleon interaction in particular, the finding of Ref.~\cite{Bai:2020yml} is a bit disturbing. We note in passing  that similar difficulties have been observed in the description of the lattice QCD simulations of the $\Lambda_c N$ interaction in the $^3S_1$-$^3D_1$ channel~\cite{Song:2021war}. In the present work, with the advent of the first accurate relativistic chiral nuclear force up to the next-to-next-to-leading order~\cite{Lu:2021gsb,Lu:2022yxb}, we revisit the lattice QCD simulations of Ref.~\cite{Inoue:2011ai} and we check whether they can be better described.

This work is organized as follows. In Sec.II, we briefly introduce the relativistic chiral nuclear force up to NNLO and explain how to extend it to study lattice QCD simulations obtained at unphysical pion masses.  In Sec.III, we contrast it with the lattice QCD $NN$ phase shifts and we show indeed that the lattice QCD data can be better described. We also highlight remaining issues to be understood in the future.  A short summary and outlook are given in Sec.IV.

\section{Theoretical framework}
The ChEFT we employ in the present work is described in detail in Refs.~\cite{Xiao:2018jot,wang:2020myr,Lu:2021gsb}. Here
we only provide a concise introduction and relevant extensions needed for studying the light quark mass~\footnote{In this work, the light quarks refer to the $u$ and $d$ quarks. We do not take into account the $s$ quark explicitly because its contribution to the nucleon-nucleon phase shifts can  be absorbed in the LECs.} dependence of the $NN$ interaction.

Up to the next-to-next-to-leading order, the covariant  chiral potential contains the following terms~\cite{Lu:2021gsb},
\begin{equation*}
V=V_{\mathrm{CT}}^{\mathrm{LO}}+V_{\mathrm{CT}}^{\mathrm{NLO}}+V_{\mathrm{OPE}}+V_{\mathrm{TPE}}^{\mathrm{NLO}}+V_{\mathrm{TPE}}^{\mathrm{NNLO}}-V_{\mathrm{IOPE}}, \text { (2) }    
\end{equation*}
where the first two terms refer to the LO $\left[\mathcal{O}\left(p^{0}\right)\right]$ and NLO $\left.\mathcal{O}\left(p^{2}\right)\right]$ contact contributions, while the next three terms denote the one-pion exchange (OPE), leading, and subleading TPE contributions. The last term represents the iterated OPE contribution. 

To study the lattice QCD data~\cite{Inoue:2011ai}, we need to extend the above chiral potential to include explicitly the pion mass dependence. Note that in the covariant power counting~\cite{Xiao:2018jot}, the pion mass is counted as of order $\mathcal{O}(p^1)$, the same as the three momentum of the nucleon. As a result, up to the next-to-next-to-leading order, we only need to explicitly consider quark mass dependence in the leading order contact and OPE contributions.

The contact potentials can be projected into different partial waves in the $|LSJ\rangle$ basis. In the present work, we are only
interested in the $^3S_1$-$^3D_1$ coupled channel. In the covaraint ChEFT, the corresponding partial wave potentials read
\begin{eqnarray}
 V_{3S1}  &=& \frac{\xi_{N}}{9}\left[C_{3S1}\left(9+R^{2}_{\pmb{p}}R^{2}_{\pmb{p'}}\right)+\hat{C}_{3S1}\left(R^{2}_{\pmb{p}}+R^{2}_{\pmb{p'}}\right)\right],
\end{eqnarray}
\begin{eqnarray}
  V_{3D1} &=& \frac{8\xi_{N}}{9}C_{3S1}R^{2}_{\pmb{p}}R^{2}_{\pmb{p'}},
\end{eqnarray}
\begin{eqnarray}
 V_{3S1-3D1}  &=& \frac{2\sqrt{2}\xi_{N}}{9}\left[C_{3S1}R^{2}_{\pmb{p}}R^{2}_{\pmb{p'}}+\hat{C}_{3S1}R^{2}_{\pmb{p}}\right],
\end{eqnarray}
\begin{eqnarray}
 V_{3D1-3S1}  &=& \frac{2\sqrt{2}\xi_{N}}{9}\left[C_{3S1}R^{2}_{\pmb{p}}R^{2}_{\pmb{p'}}+\hat{C}_{3S1}R^{2}_{\pmb{p'}}\right],
\end{eqnarray}
where $ \xi_{N}=4 \pi N_{p}^{2}N_{p}^{'2}$, $R_{\pmb{p}}=|\pmb{p}|/(E_{p}+M)$,  $R_{\pmb{p'}}=|\pmb{p'}|/(E_{p'}+M)$, $N_{P}=\sqrt{\frac{E_{P}+M}{2 M}}$, $E_{P}=\sqrt{\pmb{p}^2+M^2}$ with $M$ the nucleon mass, $\boldsymbol{p}$ and $\boldsymbol{p}^{\prime}$ are initial and final three momentum.

The leading order OPE potential reads,
\begin{equation}
\begin{aligned}
V_{O P E}\left(\boldsymbol{p}^{\prime}, \boldsymbol{p}\right) &=-\left(g_{A}^{2} / 4 f_{\pi}^{2}\right) \times\left(\bar{u}\left(\boldsymbol{p}^{\prime}, s_{1}^{\prime}\right) \boldsymbol{\tau}_{\mathbf{1}} \gamma^{\mu} \gamma_{5} q_{\mu} u\left(\boldsymbol{p}, s_{1}\right)\right) \\
& \times\left(\bar{u}\left(-\boldsymbol{p}^{\prime}, s_{2}^{\prime}\right) \boldsymbol{\tau}_{\mathbf{2}} \gamma^{\nu} \gamma_{5} q_{\nu} u\left(-\boldsymbol{p}, s_{2}\right)\right) \\
& /\left(\left(E_{p^{\prime}}-E_{p}\right)^{2}-\left(\boldsymbol{p}^{\prime}-\boldsymbol{p}\right)^{2}-m_{\pi}^{2}\right)
\end{aligned}
\end{equation}
where  $s_{1}\left(s_{1}^{\prime}\right)$, $s_{2}\left(s_{2}^{\prime}\right)$ are spin projections,  and $u(\bar{u})$ are Dirac spinors,
$u(\boldsymbol{p}, s)=N_{P}\left(\begin{array}{c}1 \\ \frac{\sigma \cdot p}{E_{P}+M}\end{array}\right) \chi_{s}$
with $\chi_{S}$ being the Pauli spinor. In addition,  $m_{\pi}$ is the pion mass, $\tau_{1,2}$ are the isospin matrices, $g_{A}=$ $1.26$, and $f_{\pi}=92.4 \mathrm{MeV}$.

To study pion mass dependent lattice QCD data,  in the covariant ChEFT, following Ref.~\cite{Bai:2020yml},    we  add a pion-mass dependent term in the  $^3S_1$-$^3D_1$ potentials. The resulting LECs then read, explicitly,
\begin{eqnarray}
  C_{3S1}  &\rightarrow & C^*_{3S1}= C_{3S1}+C_{3S1}^{\pi} m_{\pi}^{2},\\
  \hat{C}_{3S1}  &\rightarrow  &\hat{C}^*_{3S1}= \hat{C}_{3S1}+\hat{C}_{3S1}^{\pi} m_{\pi}^{2}.
\end{eqnarray}
To be consistent, we also need to take into account the quark mass dependence of the two LECs, $g_A$ and $f_\pi$. For this, we follow Ref.~\cite{Beane:2002vs,Beane:2002xf}. They read up to the next-to-leading order,
\begin{equation}\label{ap1}
  f_{\pi}(m_\pi)=f^{(0)}_{\pi}\left[1-\frac{m^{2}_{\pi}}{4\pi^2(f^{(0)}_{\pi})^2}\log\left(\frac{m_{\pi}}{m^\mathrm{PHYS}_{\pi}}\right)+\frac{m^{2}_{\pi}}{8\pi^2(f^{(0)}_{\pi})^2}\overline{l}_{4}\right]
\end{equation}
\begin{equation}\label{ap2}
  g_{A}(m_\pi)=g^{(0)}_{A}\left[1-\frac{2(g^{(0)}_{A})^2+1}{8\pi^2(f^{(0)}_{\pi})^2}m^{2}_{\pi}\log\left(\frac{m^{2}_{\pi}}{\lambda^2}\right)+\mathcal{O}(m^{2}_{\pi})\right]
\end{equation}
where $\overline{l}_{4}=4.4 \pm 0.2$, and the physical value of the pion mass, $m^\mathrm{PHYS}_{\pi} = 139$ MeV, has been used to determine the
constants in the chiral limit, $g_A^{(0)}= 1.10$ and $f_\pi^{(0)}=126.49$ MeV. We have retained only the leading chiral-logarithmic contribution to $g_{A}$, and have chosen
a renormalization scale of $\lambda = 500$ MeV.

Once the chiral potentials are fixed, we solve the relativistic Blankenbecler-Sugar (BbS) scattering equation~\cite{Blankenbecler:1965gx} to obtain
the scattering amplitudes,

\begin{eqnarray}
   T(\bm{p}',\bm{p},s) &=& V(\bm{p}',\bm{p},s)+\int\frac{\mathrm{d}^3\bm{k}}{(2\pi)^3}V(\bm{p}',\bm{k},s) \nonumber\\
   &\times& \frac{M^2}{E_k}\frac{1}{\bm{q}_{cm}^2-\bm{k}^2-i\epsilon}T(\bm{k},\bm{p},s) ,
\end{eqnarray}
where $|\bm{q}_{cm}|=\sqrt{s/4-M^2}$ is the nucleon momentum on the mass shell in the center of mass (c.m.) frame, and $\sqrt{s}$ is the total energy of the two-nucleon system in the c.m. frame.

The partial wave $S$ matrix is related to the on-shell
$T$ matrix by

\begin{align}
S_{L'L}^{SJ}({p}_{\text{cm}}) = \delta_{L'L}^{SJ} + 2 i \rho T_{L'L}^{SJ}({p}_{\text{cm}}),\quad \rho=-\frac{|{p}_{\text{cm}}|M^2}{16\pi^2E_{\text{cm}}},
\label{SRELT}
\end{align}
where the phase space factor $\rho$ is determined
by the elastic unitarity of the relativistic scattering equation. 
In order to calculate the phase shifts in the coupled
channels ($J > 0$), we use the “Stapp”- or “bar”- phase shift parametrization ~\cite{PhysRev.105.302} of the $S$ matrix, which can be written as

\begin{equation}
  S=\begin{pmatrix}
\cos2\varepsilon_Je^{(2i\delta^{1J}_{-})}&i\sin2\varepsilon_Je^{[i(\delta^{1J}_{-}+\delta^{1J}_{+})]}\\
i\sin2\varepsilon_Je^{[i(\delta^{1J}_{-}+\delta^{1J}_{+})]}&\cos2\varepsilon_Je^{(2i\delta^{1J}_{+})}
\end{pmatrix},
\end{equation}
where the subscript  “$+$” denotes $J+1$, and  “$-$” denotes $J-1$.

\begin{figure*}[htpb]
\includegraphics[scale=0.29]{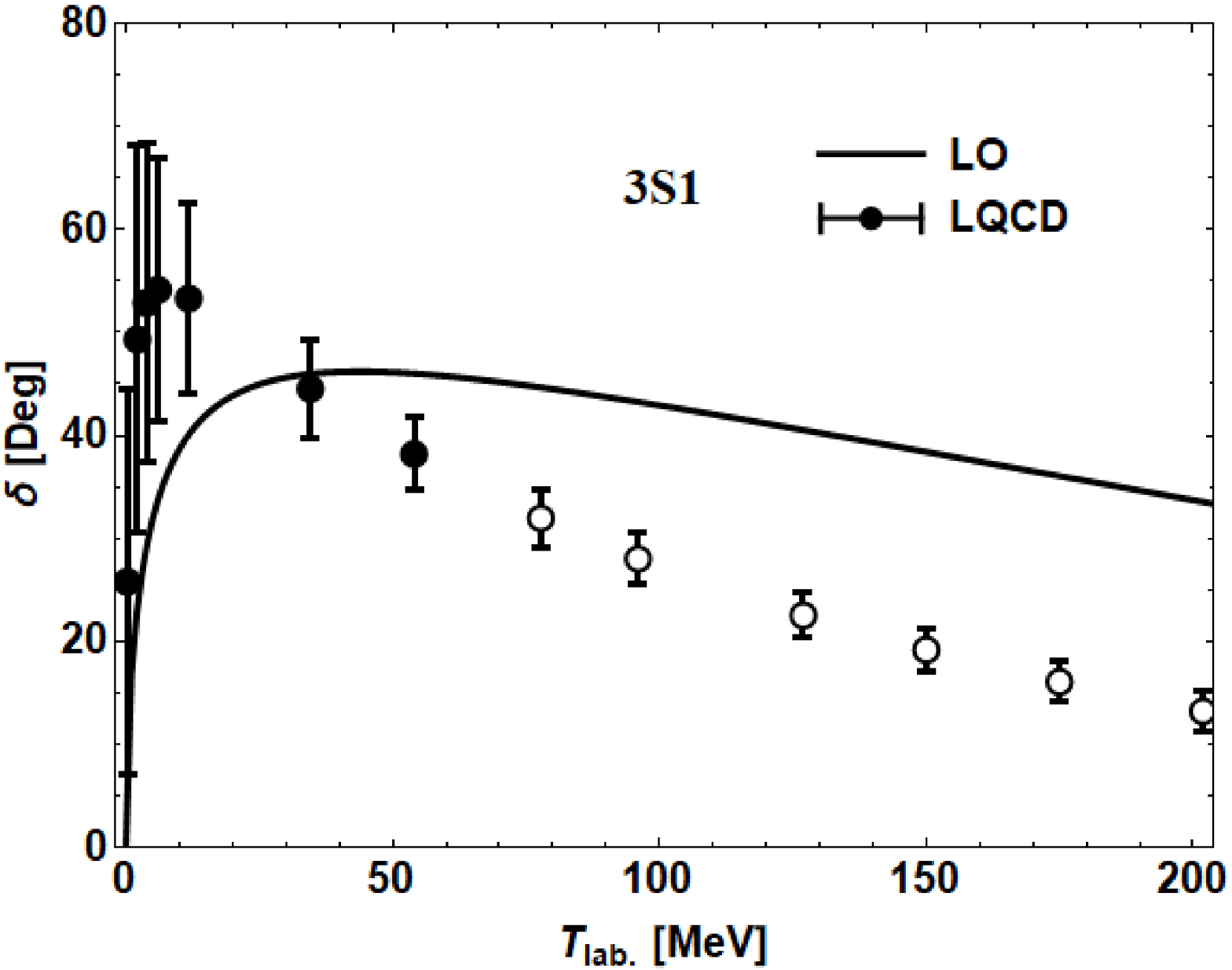}
\includegraphics[scale=0.29]{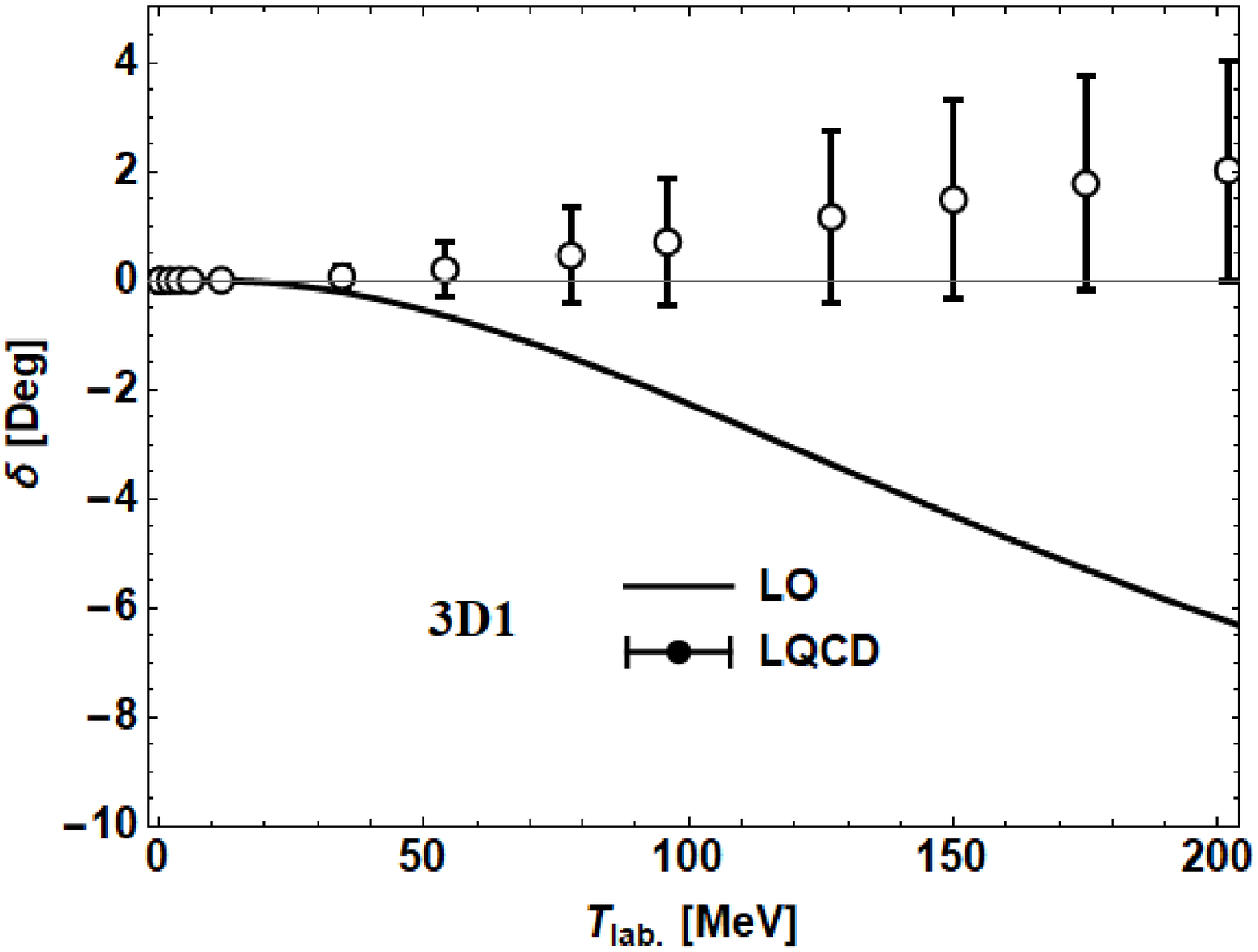}
\includegraphics[scale=0.28]{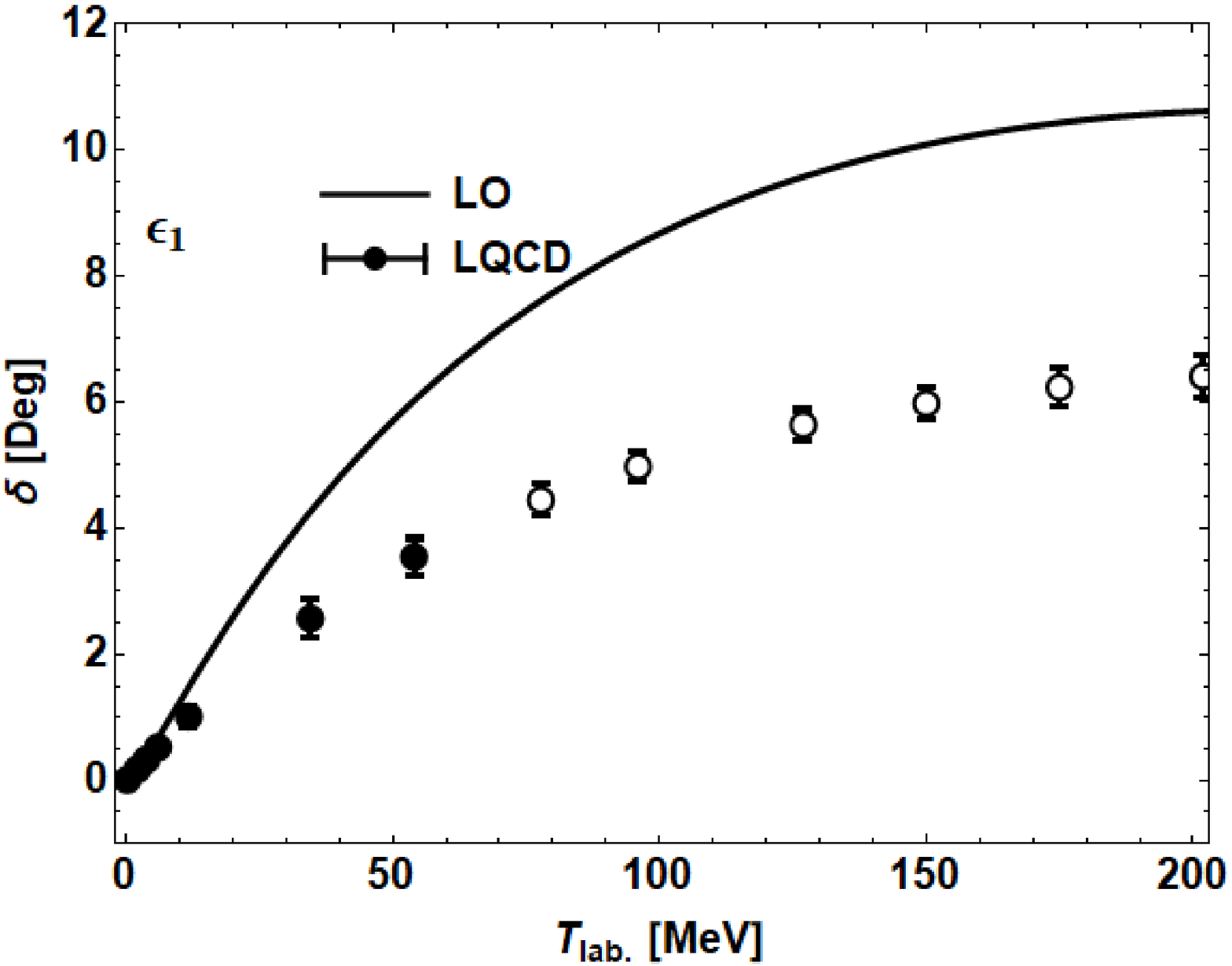}
\caption{ 
 Fitting to $\delta_{3S1}$ and $\varepsilon_1$ and  predicting $\delta_{3D1}$  in the covariant ChEFT at LO. The solid points with error bars are the lQCD469 data~\cite{Inoue:2011ai} fitted, while those empty circles are not fitted. \label{fig:LO3SD} }
\end{figure*}

\begin{table*}[htpb]
\centering
\caption{LECs for the relativistic LO results shown in Fig.~\ref{fig:LO3SD}.  Note that the cutoff $\Lambda$ is that determined in Ref.~\cite{Lu:2021gsb} by fitting to the $NN$ physical phase shifts, while  $C_{3S1}^\pi$ and $\hat{C}_{3S1}^\pi$ are fitted to the lQCD469 data~\cite{Inoue:2011ai} with the constraints that at the physical point $C_{3S1}^*$ and $\hat{C}_{3S1}^*$ reduce to their counterparts of Ref.~\cite{Lu:2021gsb}.}\label{tab:LO3SD}
\begin{tabular}{c c c  c c c}
 \hline\hline
   $C_{3S1}$  [$10^{2}$ GeV$^{-2}$]   &  $C^{\pi}_{3S1}$  [$10^{2}$ GeV$^{-4}$]  & $\hat{C}_{3S1}$   [$10^{4}$ GeV$^{-2}$]  & $\hat{C}^{\pi}_{3S1}$  [$10^{4}$ GeV$^{-4}$]  &   $\Lambda$ [GeV] & $\chi^2/\mathrm{d.o.f.}$ \\ [0.3ex]
 \hline
  0.09 & 1.00(10) & -0.17 & 0.89(42) & 0.60 & 1.25\\[0.3ex]
 \hline\hline
\end{tabular}
\end{table*}

\section{Results and discussions}
In this section, we first show that the LO covariant ChEFT cannot describe well the lQCD469 data. Then we turn to the NLO and NNLO covariant ChEFT. We perform the standard $\chi^2$ minimization using MINUIT~\cite{green1994minuit} to fix the relevant LECs by fitting to the lQCD data  at several energies as specified below.

\subsection{LO covariant ChEFT study}

\begin{figure*}[htpb]
\includegraphics[scale=0.29]{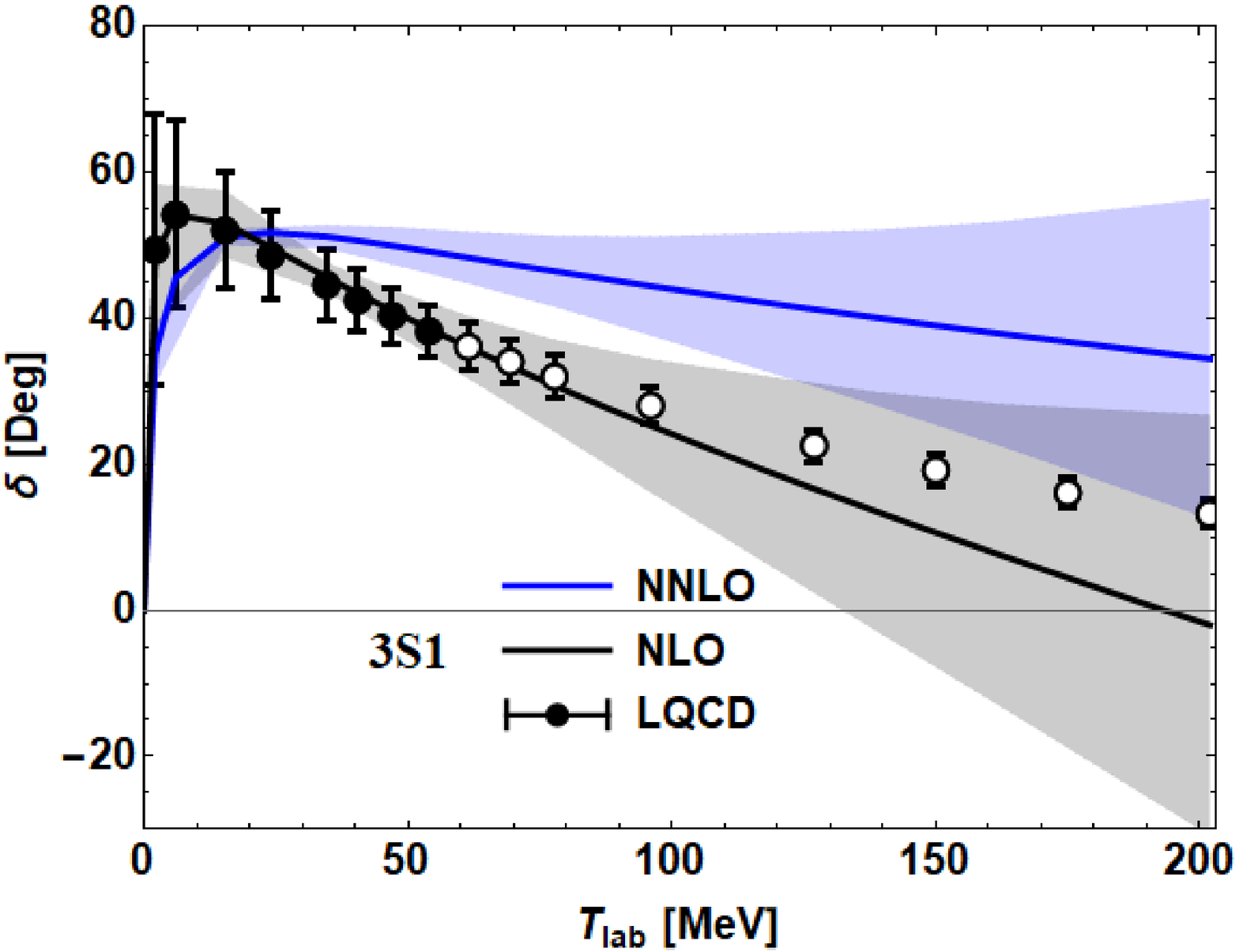}
\includegraphics[scale=0.29]{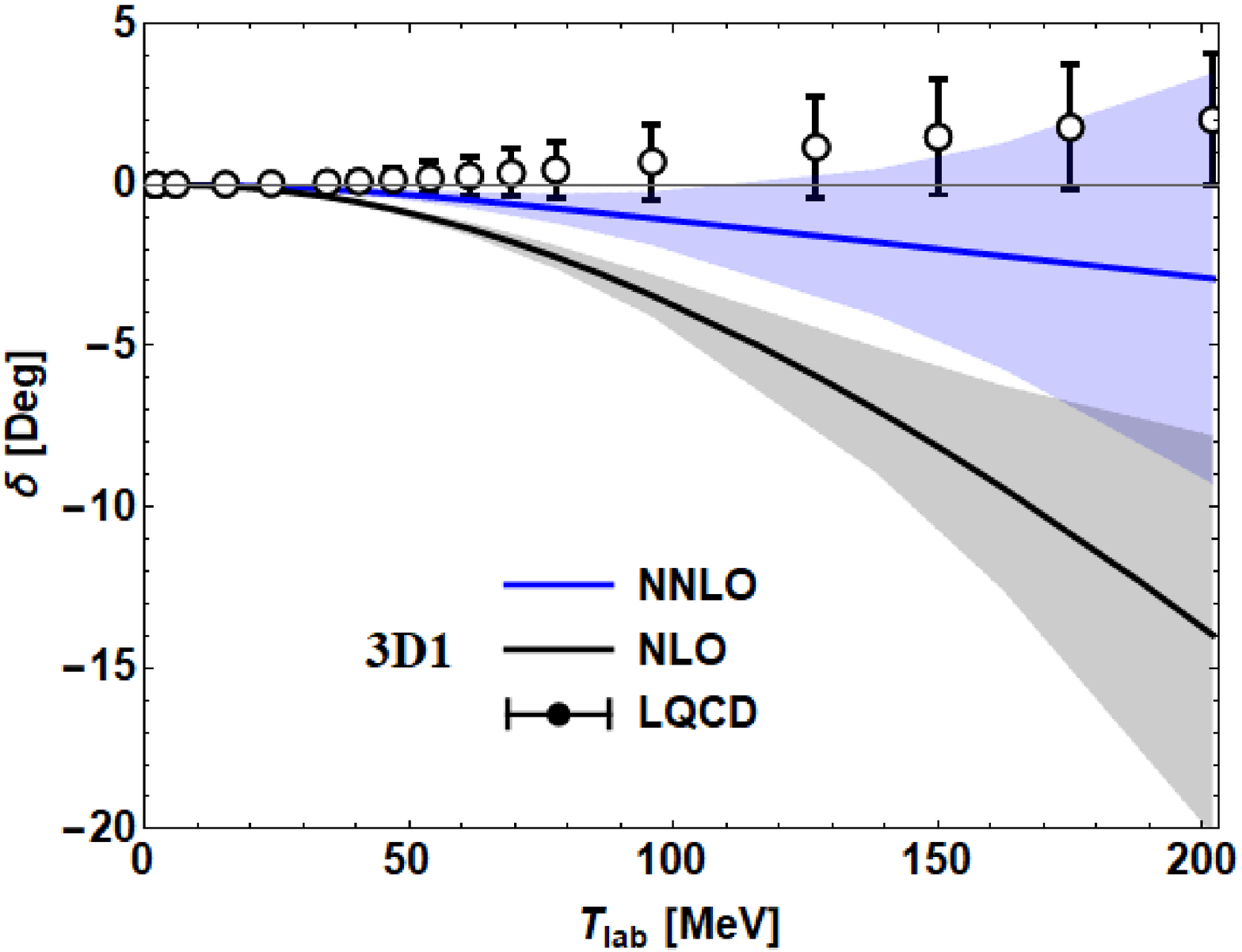}
\includegraphics[scale=0.28]{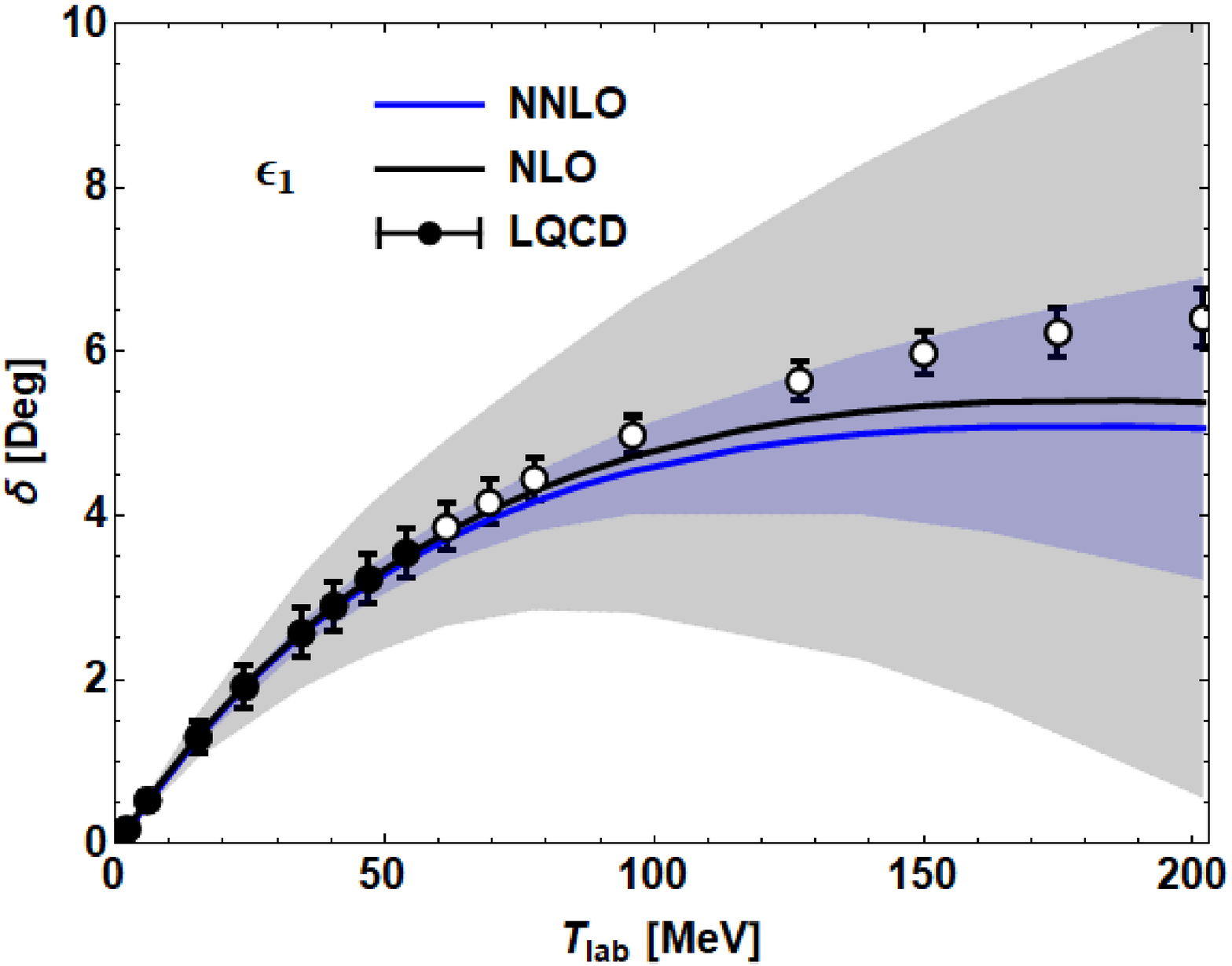}
\caption{Fitting to $\delta_{3S1}$ and $\varepsilon_1$ and predicting $\delta_{3D1}$ up to NLO and NNLO in the covariant ChEFT. The solid points with error bars are the lQCD469 data~\cite{Inoue:2011ai} fitted, while those empty circles are not fitted. The gray and blue bands represent uncertainties calculated in the Bayesian method for a DoB of 68\%. \label{fig:NLO3SD}}
\end{figure*}

\begin{table*}[htpb]
\centering
\caption{LECs for the relativistic NLO and NNLO results shown in Fig.~\ref{fig:NLO3SD}.  Note that the cutoff $\Lambda$ is that determined in Ref.~\cite{Lu:2021gsb} by fitting to the $NN$ physical phase shifts, while  $C_{3S1}^\pi$ and $\hat{C}_{3S1}^\pi$ are fitted to the lQCD469 data~\cite{Inoue:2011ai} with the constraints that at the physical point $C_{3S1}^*$ and $\hat{C}_{3S1}^*$ reduce to their counterparts  of Ref.~\cite{Lu:2021gsb}. \label{tab:NLO3SD}}
\begin{tabular}{c c c c  c c c}
 \hline\hline
   &   $C_{3S1}$[$10^{4}$ GeV$^{-2}$] & $C^{\pi}_{3S1}$ [$10^{4}$ GeV$^{-4}$]  &  $\hat{C}_{3S1}$[$10^{4}$ GeV$^{-2}$]  & $\hat{C}^{\pi}_{3S1}$ [$10^{4}$ GeV$^{-4}$]  &$\Lambda$ [GeV]& $\chi^2/\mathrm{d.o.f.}$\\ [0.3ex]
 \hline
NLO  & -5.66 & -52.73(68) & 53.04 & -24.82(2.04) & 0.60 & 0.02 \\[0.3ex]
 \hline
NNLO  & 17.52  & -9.92(1.02) & -22.36 & -27.36(3.05) & 0.90 & 1.68 \\[0.3ex]
 \hline\hline
\end{tabular}
\end{table*}
First, we fit to the lQCD469 data for $\delta_{3S1}$  and $\varepsilon_1$ with the LO covariant ChEFT up to $T_\mathrm{lab.}\approx50$ MeV, which consist of in total 14 data.~\footnote{At LO in the covariant ChEFT, the same two LECs $C_{3S1}$ and $\hat{C}_{3S1}$ contribute to $\delta_{3S1}$, $\delta_{3D1}$, and $\varepsilon_1$. Therefore, we can determine the two LECs by fitting to either one, two, or three of the coupled-channel phase shifts. However, as we show in the present work, it is not possible to reproduce $\delta_{3D1}$, therefore we choose to fit $\delta_{3S1}$ and $\varepsilon_1$.}  The cutoff is fixed at 600 MeV, which provides the best description of all the $J=0$ and $1$ partial wave phase shifts up to $T_\mathrm{lab.}=200$ MeV~\cite{Lu:2021gsb}. The so-obtained LECs and $\chi^2/\mathrm{d.o.f}$ are given in Table \ref{tab:LO3SD} and the resulting phase shifts are compared with the lattice QCD data in Fig.~\ref{fig:LO3SD}. Clearly, $\delta_{3S1}$ and $\varepsilon_1$ cannot be well described.  The same can be said  about the predicted  $\delta_{3D1}$. In particular, the lattice QCD $\delta_{3D1}$ tends to be positive while the ChEFT predictions are negative.   We note that the fittings are of similar quality as those of Ref.~\cite{Bai:2020yml}, where the physical $NN$ phase shifts are fitted simultaneously with the lQCD469 data.  

From the LECs tabulated in Table~\ref{tab:LO3SD}, we can estimate that for a pion mass of 469 MeV, the contributions of the pion-mass dependent terms are sizable (comparable to the pion-mass independent contributions).  On the other hand, as we cannot obtain an acceptable description of the lattice QCD data, we should not over-interpret the meaning of these LECs.

Clearly, although the leading order covariant Chiral EFT can describe the physical data in the $^3S_1$-$^3D_1$ coupled channel reasonably well~\cite{Ren:2016jna,Ren:2017yvw,wang:2020myr,Lu:2021gsb}, its description of the lQCD469 data is not very satisfactory. In particular, it consistently yields negative $\delta_{3D1}$ while the lattice QCD simulations prefer positive (but consistent with zero) values. In addition, both at the physical pion mass and $m_\pi=469$ MeV, the leading order covariant ChEFT also cannot describe well the mixing angle, though at a qualitative level the description is acceptable.  It is therefore intriguing to check whether at higher chiral orders, the lattice QCD data can be better described. It is also necessary to point out that in the above study, as well as in Ref.~\cite{Bai:2020yml}, the inclusion of the pion-mass dependence is not very consistent, because according to the covariant power counting~\cite{Xiao:2018jot}, the pion-mass dependent term is of order $\mathcal{O}(p^2)$, therefore in the potential one should take into account the next-to-leading contact  as well as  two-pion-exchange contributions. Now this has become feasible because both have recently been achieved in the covariant ChEFT~\cite{Xiao:2020ozd,Wang:2021kos,Lu:2021gsb}.

\subsection{NLO and NNLO covariant ChEFT study}
In this subsection, we study the lQCD469 data in covariant ChEFT up to NLO and NNLO. The number of LECs at NLO and NNLO are the same. The only difference is that at NNLO, four more TPE diagrams contribute~\cite{Xiao:2020ozd,Wang:2021kos}, which are purely predictions because they are determined by four LECs, i.e., $c_{1,2,3,4}$, which are fixed by fitting to the pion-nucleon scattering data~\cite{Chen:2012nx,Lu:2018zof}. In Ref.~\cite{Lu:2021gsb}, it was shown that although up to $T_\mathrm{lab.}=200$ MeV, the description of the physical $NN$ phase shifts up to $J\le2$ is of similar quality, but at NNLO, one can achieve a better description of higher energy data and the chiral truncation uncertainties are reduced. Therefore, we study the lQCD469 data at both chiral orders.

It is necessary to stress that in all our studies at LO, NLO, and NNLO, we have only fitted the LECs that determine the pion mass dependence of the relevant leading order LECs $C_{3S1}^\pi$ and $\hat{C}_{3S1}^\pi$. As a result, the descriptions of the physical $NN$ phase shifts at the corresponding order remain the same, see, e.g., Ref.~\cite{Lu:2021gsb}. 

 Similar to the leading order study, we fit to $\delta_{3S1}$ and $\varepsilon_1$ up to $T_\mathrm{lab.}\approx50$ MeV~\footnote{As a matter of fact, at NLO, one can fit to much higher $T_\mathrm{lab.}$, but at NNLO one cannot. Therefore, we have only fitted to the lQCD469 data up to $T_\mathrm{lab.}\approx50$ MeV.} and predict $\delta_{3D1}$. The so-obtained LECs  are given in Table~\ref{tab:NLO3SD} and the resulting phase shifts and mixing angle are compared with the lattice QCD data in Fig.~\ref{fig:NLO3SD}. The gray and blue bands are truncation uncertainties generated in the Bayesian method~\cite{Furnstahl:2015rha,Melendez:2017phj,Melendez:2019izc} for a degree of belief (DoB) of 68\%~\footnote{For a detailed discussion in the present context, see Ref.~\cite{Lu:2021gsb}.}. A few things are noteworthy. First, different from the leading order study, the NLO descriptions of the lattice QCD $\delta_{3S1}$ and $\varepsilon_1$ are quite good up to $T_\mathrm{lab.}=100$ MeV, and considering uncertainties even up to $T_\mathrm{lab.}=200$ MeV. However, the predicted $\delta_{3D1}$  is still far away from the lattice QCD data, similar to the leading order study shown in Fig.~\ref{fig:LO3SD}. On the other hand, somehow surprisingly the NNLO results are much worse than the NLO results, although within the energy region fitted, the description of the lattice QCD data is reasonable. Given the poor description of $\delta_{3S1}$  and $\varepsilon_1$  by the NNLO fit, its better prediction of $\delta_{3D1}$ compared to that of the NLO fit might be accidental and should not be over-interpreted.

It must be stressed that as both the NLO and the NNLO covariant ChEFT can describe the physical $NN$ phase shifts equally well, the deterioration of the NNLO fit to the lattice QCD data can only be traced back to the different pion mass dependence encoded in the potential. As the only difference between them is the TPE contributions induced by the pion-nucleon coupling constants $c_{1,2,3,4}$, we can reasonably speculate that their pion mass dependence might be nontrivial, contrary to what implied here that at the order of current interest, i.e., one can completely neglect their pion mass dependence. This can be checked in the future if pion-nucleon scattering can be simulated on the lattice and studied in covariant baryon chiral perturbation theory.

\section{Summary and outlook}
In this work, we studied the state-of-the-art lattice QCD simulations of the nucleon-nucleon $^3S_1$-$^3D_1$ coupled-channel phase shifts obtained at a pion mass of $469$ MeV in the covariant chiral effective field theory up to the next-to-next-to-leading order. We showed that at the next-to-leading order, the lattice QCD simulations, particularly, those of $\delta_{3S1}$ and $\varepsilon_1$, can be well described. While at the next-to-next-to-leading order, the description deteriorated, which could be naively attributed to the subleading two-pion-exchange contributions, whose low-energy constants may contain non-trivial light quark mass dependence. We have also applied the Bayesian method to provide truncation uncertainties for both the next-to-leading-order and next-to-next-to-leading-order results. Considering them, the NLO relativistic chiral nuclear force provides a satisfactory description of the lattice QCD $\delta_{3S1}$ and $\varepsilon_1$ up to $T_\mathrm{lab.}=200$ MeV.

The present study showed that although for a pion mass of 469 MeV, the covariant baryon chiral perturbation theory can still provide a good description of the lattice QCD data, the importance of the pion-mass dependent contact terms and the fact that only up to the next-to-leading order one can achieve this clearly demonstrates that for the nucleon-nucleon interaction, the chiral extrapolation is highly nontrivial. This should be kept in mind in future studies of baryon-baryon interactions on the lattice.  In addition, all the ChEFT studies preferred negative $\delta_{3D1}$, contrary to the lattice QCD simulations (at least their central values). This needs to be better understood in the future.

It should be noted that our current study is tied to the HALQCD simulations of the nucleon-nucleon interaction. It will be of great value if other lattice QCD collaborations can perform more simulations, which may help better understand the pion mass dependence of the nucleon-nucleon interaction and baryon-baryon interactions in general. In recent years, a large amount of  the so-called exotic hadrons, which cannot easily fit into the conventional quark model, have been discovered at worldwide high energy facilities, such as LHC, KEKB, and BEPC. Because many of them are located close to the thresholds of two  conventional hadrons, they are believed to be hadronic molecules. Clearly, in such a picture, lattice QCD simulations of hadron-hadron interactions are very important to understand these exotic hadrons. Our current work could also be of relevance to such studies.

\section{Acknowledgements}

This work is partly supported by the National Natural Science Foundation of China under Grants No.11735003, No.11975041,  and No.11961141004, and the fundamental Research Funds for the Central Universities. Junxu Lu acknowledges support from the National Natural Science Foundation of China under Grant No.12105006 and China Postdoctoral Science Foundation under Grant No. 2021M690008.

\clearpage


\bibliography{nn-mpi}
\end{document}